\documentclass[proceedings]{JHEP3}

\PrHEP{PrHEP hep2001}                   
\conference{International Europhysics Conference on HEP}

\usepackage{epsfig}                   

\newcommand{\afact}{{\cal A}_{\mathrm{fact}}}
\newcommand{\anon}{{\cal A}_{\mathrm{non}}}
\newcommand{\asoft}{{\cal A}_{\mathrm{non,\, soft}}}
\newcommand{\ahard}{{\cal A}_{\mathrm{non,\, hard}}}

\title{Designer mesons for exploring factorization in $b$~decays}

\author{\speaker{Markus Diehl}\\
        Institut f\"ur Theoretische Physik E, RWTH Aachen, 52056 Aachen,
        Germany\\
        E-mail: \email{mdiehl@physik.rwth-aachen.de}} 

\abstract{I explain how various aspects of factorization in exclusive
$b$ decays can be studied with mesons having a small decay constant or
spin greater than one.}

\begin{document}

\section{Testing factorization}

An outstanding task in heavy-flavor physics is to understand the
strong-interaction dynamics in exclusive decays of $b$ mesons or
baryons.  Often this is a condition \emph{sine qua non} for extracting
information on \textit{CP} violation or possible physics beyond the
standard model.  A highly successful tool for this task is the concept
of factorization \cite{Bauer:1987bm}, where a quark-antiquark pair
created in the decay of the $b$-quark forms a meson independently of
the remaining process (Fig.~\ref{fig:fact}a).  To make full use of
this tool we need to understand how well it works quantitatively, and
under which circumstances, i.e., for which decay channels.

\FIGURE{
\epsfxsize=0.8\textwidth
\epsfbox{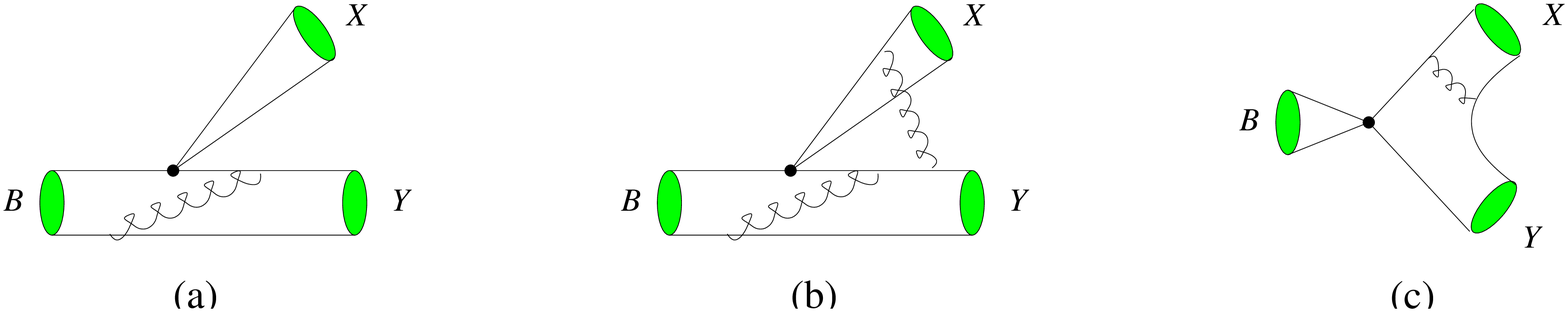} 
\caption{Example diagrams for (a) the factorization mechanism, (b)
non-factorizing gluon exchange, (c) annihilation.\label{fig:fact}} }

To be specific let us decompose a decay amplitude into its factorizing
and non-factorizing parts, ${\cal A} = \afact + \anon$.  Mechanisms
contributing to $\anon$ are for instance non-factorizing gluon
exchange (Fig.~\ref{fig:fact}b), annihilation (Fig.~\ref{fig:fact}c),
intrinsic charm in the meson wave functions \cite{Brodsky:2001yt}, or
so-called charming penguins \cite{Ciuchini:1997hb}.  In decays where
factorization works well we have $|\anon| \ll |\afact|$, and to infer
on the size of $\anon$ from the measured branching ratio is not easy.
An alternative strategy is to take channels where $\afact$ is absent
or suppressed because of some symmetry: then $\anon$ is much more
``visible''.  Information from such channels can then be used to
estimate $\anon$ in decays where one can argue that the
non-factorizing decay mechanisms contribute with similar size.

\section{Decays into designer mesons}

A wide range of mesons is almost designed to ``switch off'' the
factorizing piece of the decay amplitude
\cite{Diehl:2001xe,Harrison:1998yr}.  Several mesons have small or
zero coupling to the vector and the axial vector currents for symmetry
reasons.  An example are the scalars $a_0(980)$ and $a_0(1450)$, whose
decay constants are proportional to the tiny difference of the $u$ and
$d$ quark masses and vanish in the limit of exact isospin symmetry.
Other examples are $b_1$, $\pi(1300)$, and to a lesser degree $K_0^*$.
Unfortunately, there is no experimental information on these decay
constants, but the theory estimates in Table~\ref{tab:friends} suggest
that they might be accessible in $\tau$-decays at present or planned
facilities.  In the heavy quark sector, the decay constant of the
$\chi_{c0}$ is zero because of charge conjugation invariance, and
there is a charmed axial meson whose decay constant vanishes in the
heavy-quark limit \cite{Veseli:1996yg}.

\TABLE{\begin{tabular}{ccccc} 
\hline\hline
 $X$         & $a_0(980)$ & $a_0(1450)$ & $\pi(1300)$ & $K_0^*(1430)$ \\
 $f_X$ [MeV] & 1.1        & 0.7         & $\le 7.2$   & 42 \\
 $B(\tau\to \nu_\tau X)$ & $3.8\cdot 10^{-6}$ & $3.7\cdot 10^{-7}$ & 
               $\le 7.3\cdot 10^{-5}$ & $7.7\cdot 10^{-5}$ \\
\hline\hline
\end{tabular}
\caption{Theory estimates of decay constants as compiled in
\protect\cite{Diehl:2001xe} and the corresponding branching ratios for
$\tau\to \nu_\tau X$.  In our convention $f_\pi \approx 131$~MeV.
\label{tab:friends}} 
}

The suppression of $\afact$ for these mesons is circumvented in decays
with penguin operators involving the scalar or pseudoscalar current.
All quark-antiquark currents in the effective Hamiltonian for $b$
decays have however spin zero or one, and for mesons of higher spin
such as $X= a_2, \pi_2, \rho_3, K_2^*, D_2^*,\chi_{c2}$ we strictly
have $\afact = 0$.

To suppress the factorization mechanism one must chose the flavor
structure of a decay mode so that the designer meson $X$ has to be
emitted from the weak current and cannot pick up the spectator from
the $B$ as does meson $Y$ in Fig.~\ref{fig:fact}a.  In
Table~\ref{tab:modes} we list some of the many channels satisfying
this criterion.  To know how important non-factorizing mechanisms are
in these modes may help us understand the dynamical origins of
factorization itself, because arguments based on color transparency
\cite{Bjorken:1989kk} and arguments starting from the color structure
($1/N_c$ counting) \cite{Buras:1986xv} do not apply to the same
decays.  Note that the decay $B^+\to K^+ \chi_{c0}$, where
factorization follows from color transparency to the extent that the
$\chi_{c0}$ has a small radius, has recently been observed
\cite{Abe:2001pf}.  If the emitted meson $X$ is made from light
quarks, color transparency predicts non-factorizing interactions to be
small if the energy-mass ratio $E_X/ m_X$ is large.  This can be
tested by comparing channels with designer mesons of different mass.

\TABLE{\begin{tabular}{lcc} 
\hline\hline
decay mode & \multicolumn{2}{c}{factorization should hold according to} \\
           & color transparency & $1/N_c$ counting \\
\hline
$\bar{B}^0\to D^+ a_0^-$        & yes & yes \\
$\bar{B}^0\to D^+ D_{s2}^-$     & no  & yes \\
$B^+\to K^+ \chi_{c0}$          & yes & no \\
$\bar{B}^0\to \pi^0 D_{2}^{*0}$ & no & no \\
\hline
$\bar{B_s}\to D_s^+ a_0^-$      & yes & yes \\
$\bar{B_s}\to D_s^+ D_2^{*-}$      & no  & yes \\
$\bar{B_s}\to \eta\, \chi_{c0}$ & yes & no \\
$\bar{B_s}\to K^0 D_2^{*0}$     & no  & no \\
\hline
$\Omega_b\to \Omega_c\, a_0^-$     & yes & yes \\
$\Lambda_b\to \Lambda_c D_{s2}^-$  & no  & yes \\
$\Lambda_b\to \Lambda\, \chi_{c0}$ & yes & no \\
$\Omega_b\to \Xi^- D_2^{*0}$       & no  & no \\
\hline\hline
\end{tabular}
\caption{Selected decays into designer mesons where the factorizing
contribution is suppressed.
\label{tab:modes}} 
}

Exploring factorization in designer decays is complementary to the
classical factorization tests with modes like $\bar{B}\to D^+ \pi^-$,
$\bar{B}\to D^+ a_1^-$, etc.  To obtain meaningful constraints on
$\anon$ there one needs high precision, both in the measurement and
the theory calculation, with decay constants and form factors being
crucial ingredients.  In designer decays, where the branching ratio
gives rather direct information on $\anon$, one can relax these
requirements.  The price to pay is typically a lower branching
fraction and the requirement to handle multi-particle final states.
Angular analysis may be necessary, e.g., in order to separate
suppressed decays into $K_0^*$ or $K_2^*$ from allowed ones into
$K_1^*$ since the three states are nearly mass degenerate and decay
predominantly into $K\pi$.

In analogy to usual factorization tests \cite{Ligeti:2001dk} it is
actually not necessary to separate resonant production of designer
mesons from continuum production: the suppression of $\afact$ holds
for instance just as well for a $K_2^*$ as for $K\pi$ continuum state
with angular momentum $J=2$.

\section{Decays into heavy-light states and hard non-factorizing
interactions} 

$B$ decays into a $D$ or $D^*$ and a light meson $X$ emitted by the
weak current allow the application of powerful theory concepts.  Among
them is QCD factorization \cite{Beneke:2000ry}, where $\anon$ can be
separated into a soft part $\asoft$ that is power suppressed in
$1/m_b$ and a part $\ahard$ of order $\alpha_s$ due to hard
interactions.  The latter corresponds to diagrams as in
Fig.~\ref{fig:fact}b and can be calculated if one knows the meson
distribution amplitude $\varphi_X(u)$ describing the transition from
the $q\bar{q}$ pair to the meson $X$.  Such a mechanism evades the
suppression discussed above.  In Fig.~\ref{fig:fact}b an interaction
takes place between the creation of the $q\bar{q}$ pair at the $b$
decay vertex and its hadronization into the meson $X$.  Even if $X$
has small or zero coupling to the \emph{local} quark-antiquark current
of the $b$ decay, its distribution amplitude $\varphi_X(u)$ need not
be small since it involves the corresponding \emph{nonlocal} current.
In fact, $|\varphi_X(u)|^2$ is related to the probability that $X$
fluctuates into a current $q\bar{q}$ pair, and we used this relation
in \cite{Diehl:2001xe} to estimate the size of the meson distribution
amplitudes.  Experimental constraints on these important quantities
could be obtained in the process $e^+ e^- \to e^+ e^- X$ when one of
the lepton beams receives a large invariant momentum transfer $Q^2$.
This has already been exploited for the mesons $\pi$, $\eta$ and
$\eta'$ \cite{Gronberg:1998fj}.

In Table~\ref{tab:branching} we estimate branching ratios for some
decay modes, both in naive and in QCD factorization.  We find
non-factorizing contributions $\ahard$ of similar size for decays into
designer mesons and for modes like $\bar{B}\to D^+ \pi^-$, where they
only amount to a few percent of the large amplitude $\afact$.  In
designer channels, on the other hand, $\ahard$ can be comparable to or
bigger than $\afact$, as comparison of the naive and QCD factorization
results in Table~\ref{tab:branching} shows.  To be sure, our rate
estimates are fraught with uncertainties from the unknown decay
constants and distribution amplitudes, and also with a strong
renormalization scale dependence of $\ahard$ at leading order in
$\alpha_s$.  Most important is however that $\ahard$ is \emph{tiny} on
the scale of, say, the amplitude for $\bar{B}\to D^+ \pi^-$ and may
well be overshadowed by the soft factorization breaking described by
$\asoft$.  Whether this is the case could be revealed by data on the
branching ratios for designer channels.  Since calculating $\asoft$ is
extremely hard for theory, this would be valuable information indeed.
Note that for some channels even the small rates we estimated should
be within current experimental reach, and measurement would be even
easier if $\asoft$ were large compared with our estimates of $\afact +
\ahard$.

\TABLE{\begin{tabular}{lccc} 
\hline\hline
decay mode & naive factorization & \multicolumn{2}{c}{QCD factorization}
\\ 
 & & ~~~$\mu=m_b$~~~ 
 & ~~~$\mu=\frac{1}{2}\rule[-1.1ex]{0pt}{1ex}m_b$~~~ \\ \hline
$\bar{B}^0 \to D^+ a_0(980)$ & $1.1\cdot 10^{-6}$
                             & $2.0\cdot 10^{-6}$ & $4.0\cdot 10^{-6}$
\\
$\bar{B}^0 \to D^+ a_0(1450)$ & $8.6\cdot 10^{-8}$
                              & $5.8\cdot 10^{-7}$ & $2.1\cdot 10^{-6}$
\\
$\bar{B}^0 \to D^+ a_2$ & 0
                              & $3.5 \cdot 10^{-7}$ & $1.7 \cdot 10^{-6}$
\\
$\bar{B}^0 \to D^+ \pi(1300)$ & $9.1\cdot 10^{-6}$
                              & $9.3\cdot 10^{-6}$ & $9.6\cdot 10^{-6}$
\\
$\bar{B}^0 \to D^+\pi_2$ & 0
                              & $1.4 \cdot 10^{-9}$ & $8.1 \cdot 10^{-9}$
\\
$\bar{B}^0 \to D^+ K^*_0(1430)$ & $2.0\cdot 10^{-5}$ & $2.0\cdot 10^{-5}$
                                & $2.1\cdot 10^{-5}$
\\
$\bar{B}^0 \to D^+ K^*_2$ & 0 & $1.9\cdot 10^{-8}$ & $9.2\cdot 10^{-8}$
\\ 
\hline\hline
\end{tabular}
\caption{Branching ratio estimates in naive factorization and in QCD
factorization to $O(\alpha_s)$ with two choices for the factorization
scale $\mu$.  We find similar values for the corresponding decays
$\bar{B}^0\to D^{*+} X^-$, $\bar{B}_s^{\protect\phantom{0}}\to D_s^{+}
X^-$, and $\bar{B}_s^{\protect\phantom{0}}\to D_s^{*+} X^-$.
\label{tab:branching}} 
}

\section{Decays into light-light states and penguins}

$B$ decays into two light mesons present a much greater complexity in
the electroweak and the strong dynamics.  Among the questions
currently under debate is the importance of annihilation graphs
(Fig.~\ref{fig:fact}c) with penguin operators, which could have a
strong impact on the study of \textit{CP} violation.  Whether
annihilation graphs can be reliably calculated is controversial:
whereas in QCD factorization they are $1/m_b$ corrections and can only
be estimated \cite{Beneke:2001ev}, their evaluation for $B \to K \pi$
in the pQCD approach of Li et al.~\cite{Keum:2001wi} found them to be
substantial and with a large phase relative to the factorizing
contribution.  Notice that depending on that phase one can obtain the
same branching fraction with a small or a large contribution from
$\anon$, see Fig.~\ref{fig:phases}a.  Hints concerning the two
scenarios sketched there could be obtained in the designer modes
$\bar{B}\to \pi^+ K_2^{*-}$ and $B^-\to \pi^- K_2^{*0}$.  Compared
with $B\to \pi K$ the factorizing amplitude $\afact$ would be
suppressed but not $\anon$, where penguin annihilation contributes,
and the total amplitude would be quite different in the two cases, see
Fig.~\ref{fig:phases}b.

\FIGURE{
\epsfxsize=0.95\textwidth
\epsfbox{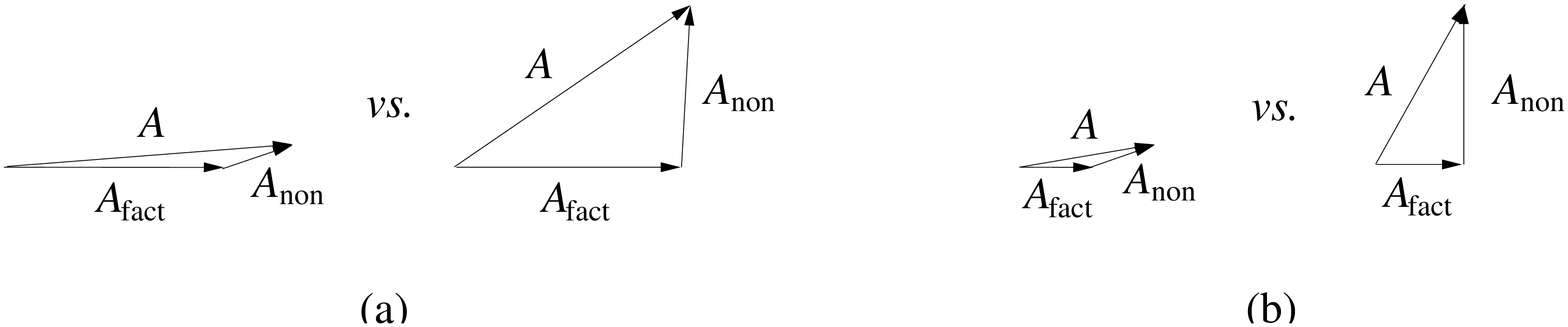} 
\caption{(a) The same $|{\cal A}|$ can be obtained with largely
different values of $\anon$, depending on its phase.  (b) Suppressing
$\afact$ but not $\anon$ leads to distinct results for $|{\cal A}|$ in
the two scenarios.
\label{fig:phases}}
}

\section{More use for designer mesons}

Designer mesons are a versatile tool to suppress selected
contributions in $b$ decay processes.  This can be used to explore the
dynamics of factorization, but there are other possibilities.  One
example is to extract the CKM phase $2\beta+\gamma$ using the
interference between mixing and decay in $B^0/ \bar{B}^0\to D^\pm
a_0^\mp$ or similar channels \cite{Diehl:2001ey}.  In contrast to the
well-studied case of $B^0/ \bar{B}^0\to D^\pm \pi^\mp$, the designer
channels can have large interference effects, the price to pay being a
lower total event rate.  Ways to investigate \textit{CP} violation in
decays to light-light final states with designer mesons have been
proposed in \cite{Laplace:2001qe}.  With the experimental
possibilities at present and in the near future, decays into designer
mesons should provide various possibilities to study important aspects
of $B$ physics.

\end{document}